\documentclass[11pt]{article}
\usepackage{fullpage}
\usepackage{amsmath}
\usepackage{amssymb}
\usepackage{hyperref}
\usepackage{amsthm}
\begin{document}

\newtheorem{theorem}{Theorem}[section]
\newtheorem{prob}[theorem]{Problem}
\newtheorem{example}[theorem]{Example}
\newtheorem{prop}[theorem]{Proposition}
\newtheorem{lemma}[theorem]{Lemma}
\newtheorem{corollary}[theorem]{Corollary}
\newtheorem{remark}[theorem]{Remark}
\newcommand{\reals}{\ensuremath{\mathbb{R}}}
\newcommand{\card}[1]{\lvert#1\rvert}
\newcommand{\comment}[1]{*\marginpar{\begin{footnotesize}*#1\end{footnotesize}}}
\newcommand{\floor}[1]{\left\lfloor#1\right\rfloor}
\newcommand{\goesto}{\rightarrow}
\newcommand{\RR}{\mathbb{R}}
\newcommand{\NN}{\mathbb{N}}
\newcommand{\suchthat}{\mathrel{:}}
\newcommand{\giventhat}{\mid}
\newcommand{\additionaldetails}[1]{}

\def\eps{\varepsilon}
\def\R{\reals}
\newcommand{\E}{\operatorname{\sf E}}
\renewcommand{\Pr}{\operatorname{\sf P}}
\def\vol{\hbox{\rm vol}}
\def\T{^{\sf T}}
\def\LL{{\cal L}}

\long\def\privremark#1{}

\title{Expanders via Random Spanning Trees}
\author{Navin Goyal \and Luis Rademacher \and Santosh Vempala}

\date{}
\maketitle

\begin{abstract}
Motivated by the problem of routing reliably and scalably in a
graph, we introduce the notion of a {\em splicer}, the union of
spanning trees of a graph. We prove that for any bounded-degree
$n$-vertex graph, the union of two {\em random} spanning trees
approximates the expansion of every cut of the graph to within a
factor of $O(\log n)$. For the random graph $G_{n,p}$, for $p> c\log{n}/n$,
two spanning trees give an expander.
This is suggested by the
case of the complete graph, where we prove that two
random spanning trees give an expander. The construction of the
splicer is elementary --- each spanning tree can be produced
independently using an algorithm by Aldous and Broder: a random walk in the graph
with edges leading to previously unvisited vertices included in the
tree.

A second important application of splicers is to graph
sparsification where the goal is to approximate every cut (and more
generally the quadratic form of the Laplacian) using only a small
subgraph of the original graph. Benczur-Karger \cite{BK} as well as
Spielman-Srivastava \cite{SS} have shown sparsifiers with $O(n \log
n/\eps^2)$ edges that achieve approximation within factors $1+\eps$
and $1-\eps$.  Their
methods, based on independent sampling of edges, need $\Omega(n\log
n)$ edges to get any approximation (else the subgraph could be
disconnected) and leave open the question of linear-size
sparsifiers. Splicers address this question for random graphs by
providing sparsifiers of size $O(n)$ that approximate every cut to
within a factor of $O(\log n)$.
\end{abstract}

\section{Introduction}
In this paper, we present a new method for obtaining sparse expanders
from spanning trees. This appears to have some interesting
consequences. We begin with some motivation.

Recovery from failures is considered one of the most important
problems with the internet today and is at or near the top of
wish-lists for a future internet. In his 2007 FCRC plenary lecture,
Shenker desires a network where ``even right after failure, routing
finds path to destination'' \cite{Shenker07}. How should routing
proceed in the presence of link or node failures?

At a high-level, to recover from failures, the network should have
many alternative paths, a property sometimes called {\em path
diversity}, which is measured by several parameters, including robustness
in the presence of failures and congestion. It is well-known that expander
graphs have low congestion and remain connected even after
many (random) failures. Indeed, there is a large literature on
routing to minimize congestion and on finding disjoint paths that is
closely related to expansion (or more generally, conductance); e.g.
\cite{RaghavanT, LeightonR, AlonC}.

However, in practice, efficient routing also needs to be compact and
{\em scalable}; in particular, the memory overhead as the network grows
should be linear or sublinear in the number of vertices. This
requirement is satisfied by routing on trees, one tree per destination.
In fact, the most
commonly used method in practice is shortest path routing which is effectively one tree per destination\footnote{It is called Open Shortest
Path First (OSPF) in networking terminology.}. Since the final destination
determines the next edge to be used,
this gives an $O(n)$ bound on the size of the
routing table that needs to be stored at each vertex. If
a constant-factor stretch is allowed, this can be reduced. For example,
with stretch $3$, tables of size $O(\sqrt{n})$ suffice as shown by
Abraham et al \cite{AGMNT04}.

The main problem with shortest-path routing or any tree-based scheme
is the lack of path diversity. Failing any edge disconnects some
pairs of vertices. Recovery is usually achieved by recomputing
shortest path trees in the remaining network, an expensive
procedure. Further, congestion can be high in principle. This is
despite the fact that the underlying graph might have high expansion, implying that
low congestion and high fault-tolerance are possible. There is some evidence that
AS-level internet topologies are expanders and some stochastic
models for networks lead to expanders \cite{MihPap2006}. However,
known algorithms that achieve near-optimal congestion use arbitrary
paths in the network and therefore violate the scalability
requirement. This raises the following question:
is it possible to have a routing scheme that is both scalable and achieves congestion and
fault-tolerance approaching that of the {\em underlying graph}?

Our work is inspired by Motiwala et al. \cite{MFV07, MEFV08},
who consider a conceptually simple extension of tree-based routing,
using multiple trees.
With one tree there is a unique
path between any two points. With two trees, by allowing a path to
switch between the trees multiple times, there could be a large
number of available paths. Motiwala et al. showed experimentally that a small number of randomly perturbed shortest path trees for each destination leads to a highly reliable routing method: the union of these trees has reliability approaching that of the underlying graph.

This raises the question whether the results of this experiment can be true in general.
I.e., for a given graph does there exist a small collection of spanning trees such that the reliability of the union approaches
that of the base graph? As a preliminary step, we study the question of whether
for a given graph the union of a few spanning trees
captures the expansion of the original graph.
Here we propose a construction that uses only a small number of trees {\em in total}
(as opposed to one tree per destination) and
works for graphs with bounded degree and for random graphs.
The trees are chosen independently from the uniform
distribution over all spanning trees, a distribution that can be
sampled efficiently with simple algorithms. The simplest of these,
due to Aldous \cite{Aldous} and Broder \cite{Broder}, is to take a random walk in the graph,
and include in the tree every edge that goes to a previously
unvisited vertex. Roughly speaking, our main result is that for bounded degree graphs
and for random graphs a small number of such trees give a subgraph with expansion comparable
to the original graph for each cut.

A second important application of splicers is to graph
sparsification where the goal is to approximate every cut (and more
generally the Laplacian quadratic form) using only a small subgraph
of the original graph. Benczur-Karger \cite{BK} as well as
Spielman-Srivastava \cite{SS} have shown sparsifiers with $O(n \log
n/\eps^2)$ edges that achieve a $1+\eps$ approximation. Their
methods are based on independent sampling of edges with carefully
chosen edge probabilities and require $\Omega(n\log n)$ edges to get
any approximation; with fewer edges, the subgraph obtained could be
disconnected. They leave open the question of the existence of
linear-size sparsifiers. Splicers, constructed using random spanning
trees, provide sparsifiers of size $O(n)$ for random graphs: When
the base graph is random, with high probability, the union of two
spanning trees approximates all cuts to
within a factor of $O(\log n)$. We state this precisely in the next
section.

\subsection{Our results}

A $k$-splicer is the union of $k$ spanning trees of a graph.
 By a
random $k$-splicer we mean the union of $k$ uniformly randomly chosen
spanning trees.  We show
that for any bounded degree graph, the union of two random spanning
trees of the graph approximates the expansion of every cut of the
graph. Using more trees gives a better approximation.  In the
following $\delta_G(A)$ stands for the set of edges in graph $G$ that
have one endpoint in $A$, a subset of vertices of $G$.

\begin{theorem} \label{thm:constant:positive}
For a $d$-regular graph $G=(V,E)$, let $U^k_G$ be a random
$k$-splicer, obtained by the union of $k$ uniformly random spanning
trees. Also let $\alpha > 0$ be a constant and $\alpha (k-1) \geq
9d^2$. Then with probability $1-o(1)$, for every $A \subset V$, we
have
$$|\delta_{U^k_G}(A)| \geq \frac{1}{\alpha \log{n}} \cdot |\delta_G(A)|.$$
\end{theorem}
Our proof of this makes novel use of a known property of random
spanning trees of a graph, namely the events of an edge in the graph
being included in the tree are negatively correlated.

Next we give a lower bound, showing that the factor $1/\log n$ is
the best possible for $k$-splicers constructed from random spanning
trees for any constant $k$.

\begin{theorem} \label{thm:constant:negative}
For every $n$, there is a bounded-degree edge expander $G$ on $n$
vertices such that with probability $1-o(1)$ the edge expansion of a
random $k$-splicer $U^k_G$ is at most $ k^2/C\log{n}$ for any $k
\geq 1$.
\end{theorem}


For the complete graph, one can do better, requiring only two
trees to get a constant-factor approximation.

\begin{theorem} \label{thm:completegraph}
The union of two uniformly random spanning trees of the complete graph on $n$ vertices
has constant vertex expansion with probability $1-o(1)$.
\end{theorem}

Since constant vertex expansion implies constant edge expansion, we
get that the union of two uniformly random spanning trees has
constant edge expansion with high probability.

Next we turn to the random graph $G_{n,p}$. Our main result here is
that w.h.p., $G_{n,p}$ has two spanning trees whose
union has constant vertex expansion. We give a simple random process
(called {\em Process $B_p$} henceforth) to find these trees.

\begin{theorem}\label{thm:expander_random}
There exists an absolute constant $C$, such that for $p \ge C\log
n/n$,  with probability $1-o(1)$, the union of two random spanning
trees from Process $B_p$ applied to a random graph $H$ drawn from
$G_{n,p}$ has constant vertex expansion.
\end{theorem}

The proof of this theorem is via a coupling lemma (Lemma
\ref{lem:coupling}) showing that a tree generated by Process $B_p$
applied to a random graph $H$ is nearly uniform among spanning trees
of the complete graph.

Theorem \ref{thm:expander_random} relates to the work of
\cite{BK,SS} and leads to the first linear-size sparsifier with
nontrivial approximation guarantees for random graphs:
\begin{theorem}\label{thm:sparsifier_gnp}
Let $p \geq C\log{n}/n$ for a sufficiently large constant $C$.  Let H be a
$G(n,p)$ random graph, and let $H'$ be the $2$-splicer obtained from it via
process $B_p$, with a weight of $pn$ on every edge.  Then with probability
$1-o(1)$, for every $A \subset V$ we have
$$ c_1 |\delta_{H}(A)|  \leq  w(\delta_{H'}(A)) \leq c_2 |\delta_H(A)| \log{n},$$
where $c_1, c_2 > 0$ are constants.
\end{theorem}
Here $w(\cdot)$ denotes the sum of the weights.

\subsection{Related work}

The idea of using multiple routing trees and switching between them
is inspired by the work of \cite{MFV07} who proposed a multi-path
extension to standard tree-based routing. The method, called {\em
Path Splicing}, computes multiple trees to each destination vertex,
using simple methods to generate the trees; in one variant, each
tree is a shortest path tree computed on a randomly perturbed set of
edge weights. Path splicing appears to do extremely well in
simulations, approaching the reliability of the underlying graph
using only a small number of trees\footnote{It has several other
features from a practical viewpoint, such as allowing end vertices
to specify paths, that we do not discuss in detail here.}.

Sampling for approximating graph cuts was introduced by Karger,
first for global min-cuts and then extended to min $s$-$t$ cuts and
flows. The most recent version due to Benczur and Karger~\cite{BK}
approximates the weight of
every cut of the graph within factors of $1+\eps$ and $1-\eps$ using
$O(n\log n/\eps^2)$ samples; edges are sampled independently with probability
inversely proportional to a connectivity parameter and each chosen
edge is weighted with the reciprocal of their probability. Recently,
Spielman and Srivastava~\cite{SS}, gave a similar method where edges
are sampled independently with probability proportional the graph
resistance and weighted in a similar way, by the reciprocal of the
probability with which they are chosen. They show that every
quadratic form of the Laplacian of the original graph is
approximated within factors $1-\eps$ and $1+\eps$. The similarity in the two
methods extends to their analysis also --- both parameters, edge
strength and edge resistance share a number of useful properties.

It has long been known that the union of three random perfect
matchings in a complete graph with even number of vertices (see, e.g.,
\cite{Goldreich}) is an expander with high probability.  Our result on the union of random spanning trees
from the complete graph can be considered as a result in a similar vein,
and our proof has a similar high-level outline.  Still, the
spanning trees case seems to be different and requires some new
ideas.

On the other hand, our result for the union of spanning trees
of bounded degree graphs doesn't seem to have any analog for the union
of matchings.  Indeed, generating random perfect matchings of graphs
is a highly nontrivial problem, the case of computing the permanent of
0--1 matrices being the special case for bipartite
graphs~\cite{permanent}.


\section{Preliminaries}

Let $G=(V,E)$ be an undirected graph. For $v \subseteq V$, define
$\Gamma(v):=\{u \in V \suchthat (u,v) \in E\}$, the set of neighbors
of $v$.  For $A \subseteq V$, define $\Gamma(A) := \cup_{v \in A}
\Gamma(v)$, and $\Gamma'(A) := \Gamma(A) \setminus A$. Finally, let
$\delta_G(A) := \{(u,v) \in E \suchthat u \in A, v \notin A\}$. The
edge expansion of $G$ is
\[
\min_{A \subseteq V, 1 \le |A|\le \card{V}/2 } \frac{|\delta_G(A)|}{|A|}.
\]
The vertex expansion of $G$ is
\[
\min_{A \subseteq V, 1 \le |A|\le \card{V}/2} \frac{|\Gamma'(A)|}{|A|}.
\]

We say that a family of graphs is an \emph{edge (resp., vertex)
expander (family)} if the edge (resp., vertex) expansion of the
family is bounded below by a positive constant.

Let $K_n$ denote the complete graph on $n$ vertices.

For $a \in \RR$, let $[a] = \{ i \in \NN \suchthat 1 \leq i \leq
a\}$.  On several occasions we will use the inequality $\binom{n}{k}
\leq (\frac{ne}{k})^k$.

\section{Uniform random spanning trees}\label{sec:negative_correlation}

Uniformly random spanning trees of graphs are fairly well-studied
objects; see, e.g., \cite{LyonsPeres}. In this section we describe
properties of random spanning trees that will be useful for us.
There are several algorithms known for generating a uniformly random
spanning tree of a graph, e.g., \cite{Aldous, Broder, Wilson,
LyonsPeres}. The algorithm due to Aldous and Broder is very
simple and will be useful in our analysis: Start a uniform random
walk at some arbitrary vertex of the graph, and when the walk visits
a vertex for the first time, include the edge used to reach that
vertex in the tree. When all the vertices have been visited we have
a spanning tree which is uniformly random \emph{regardless of the
initial vertex}.

A well-known fact (e.g.~\cite{Lovasz}) about uniform random spanning
trees is that the probability that an edge $e$ is chosen in a
uniform random spanning tree, is equal to the effective resistance
of $e$: Let each edge have unit resistance, then the effective
resistance of $e$ is the potential difference applied to the
endpoints of $e$ to induce a unit current. This fact shows a
connection of our work with \cite{SS}, who sample edges in a graph
according to their effective resistances to construct a sparsifier.

For a connected base graph $G=(V,E)$, random variable $T_G$ denotes
a uniformly random spanning tree of $G$. $U^k_G$ will denote the
union of $k$ such trees chosen independently. For edge $e \in E$,
abusing notation a little, we will refer to events $e \in E(T_G)$
and $e \in E(U^k_G)$ as $e \in T_G$ and $e \in U^k_G$.

\paragraph{Negative correlation of edges.} The events
of various edges belonging to the random spanning tree are
negatively correlated: For any subset of edges $e_1, \ldots, e_k \in
E$ we have
\begin{align} \label{eqn:negative_correlation}
\Pr[e_1, e_2, \ldots, e_k \in T_G] \:\: \leq \:\: \Pr[e_1 \in T_G] \Pr[e_2 \in T_G] \dotsm \Pr[e_k \in T_G].
\end{align}
A similar property holds for the complementary events:
\begin{align} \label{eqn:negative_correlation_complement}
\Pr[e_1 \notin T_G, \ldots, e_k \notin T_G] \:\: \leq \:\: \Pr[e_1 \notin T_G] \Pr[e_2 \notin T_G] \dotsm \Pr[e_k \notin T_G].
\end{align}
These are easy corollaries of \cite[Theorem 4.5]{LyonsPeres}, 
which in turn is based on the work of Feder and Mihail \cite{FM}.

\paragraph{Negatively correlated random variables and tail bounds.}
For $e \in E$, define indicator random variables $X_e$ to be $1$ if
$e \in T$, and $0$ otherwise. Then we can rewrite
\eqref{eqn:negative_correlation} as follows.

For any subset of edges $e_1, \ldots, e_k \in E$ we have
\begin{align}\label{negcorr}
\E[X_{e_1} \dotsm X_{e_k}] \leq \E[X_{e_1}] \dotsm \E[X_{e_k}].
\end{align}

For random variables $\{X_e\}$ satisfying \eqref{negcorr} we say
that $\{X_e\}$ are \emph{negatively correlated}.  Several closely
related notions exist; see Dubhashi and
Ranjan~\cite{DubhashiRanjan98}, and Pemantle~\cite{Pemantle}.
\cite{DubhashiRanjan98} gave a property of negative correlation that
will be useful for us: It essentially says that Chernoff's bound for
the tail probability for sums of independent random variables
applies unaltered to negatively correlated random variables. More
precisely, we will use the following version of Chernoff's bound.
\begin{theorem} \label{thm:Chernoff_negative}
Let $\{X_i\}_{i=1}^n$ be a family of 0--1 negatively correlated random variables
such that $\{1-X_i\}_{i=1}^n$ are also negatively correlated. Let $p_i$ be the
probability that $X_i=1$.
Let $p:= \frac{1}{n} \sum_{i \in [n]} p_i$.
Then for $\lambda > 0$
$$ \Pr[\sum_{i \in [n]} X_i < pn - \lambda] \leq e^{-\lambda^2/(2pn)}. $$
\end{theorem}

\begin{proof}
The proof splits into two steps: In the first step we prove that for
arbitrary $\lambda$ we have
\begin{align} \label{moment}
\E[\exp(\lambda \sum_{i=1}^{n} X_i)] \leq \prod_{i=1}^n \E[\exp(\lambda X_i)].
\end{align}
The second step is a standard Chernoff bound argument as in the proof of Theorem A.1.13
in \cite{AlonSpencer}.  Since
the first step is not well-known and is not hard, we provide a proof
here.  In this, we basically follow Dubhashi and
Ranjan~\cite{DubhashiRanjan98}.

The case $\lambda = 0$ is trivially true. We now prove \eqref{moment} for
$\lambda > 0$. Since $X_i$'s take $0$--$1$ values, for
any integers $a_1, \ldots, a_n > 0$, we have $X_1^{a_1} X_2^{a_2}
\dotsm X_n^{a_n} = X_1 X_2 \dotsm X_n$.  Now, writing $\exp(\lambda
\sum_{i=1}^{n} X_i)$ using the Taylor series for $e^x$, and
expanding each summand, we get a sum over various monomials over the
$X_i$'s.  For each monomial we have by the definition of negative
correlation that $\E[X_1 \dotsm X_n] \leq \prod_{i=1}^{n} \E[X_i]$.
This gives \eqref{moment} for $\lambda >0$.

For $\lambda <0$, a similar argument using $1-X_i$ in the
role of $X_i$ gives \eqref{moment}.
\end{proof}

\section{Expansion when base graph is a complete graph}

Our proof here has the same high-level outline as the proof for
showing that the union of three random perfect matchings in a
complete graph with even number of vertices is a vertex-expander
(see, e.g., \cite{Goldreich}): One shows that for any given vertex
set $A$ of size $\leq n/2$, the probability is very small for the
event that $\card{\Gamma'(A)}$ is small in the union of the
matchings. A union bound argument then shows that the probability is
small for the existence of any set $A$ with $\card{\Gamma'(A)}$
small. However, new ideas are needed because spanning trees are
generated by the random walk process, which appears to be more
complex to analyze than random matchings in complete graphs.

\begin{proof}[Proof (of Theorem~\ref{thm:completegraph}).]
For a random spanning tree $T$ in $K_n$ and given $A \subseteq V$, $\card{A} =
a$, we will give an upper bound on the probability that
$\card{\Gamma'_T(A)} \leq ca$, for a given expansion constant $c$
(Recall that $\Gamma'_T(A)$ denotes the set of vertices in $V \setminus A$
that are neighbors of vertices in $A$ in the graph $T$). To this end, we will fix a set
$A' \subseteq V \setminus A$ of size $\floor{ca}$ and
we will bound the probability that $\Gamma'_T(A) \subseteq A'$, and, to conclude,
use a union bound over all possible choices of $A$ and $A'$.
Without loss of generality the vertices are labeled
$V = \{1, \dotsc, n\}$, $A = [a] = \{1, \dotsc, a \}$ and $A' = \{a+1, \floor{ca}\}$.
More precisely, the union bound is the following: the probability that there exists
a set $A \subseteq V$ such that $\card{A} \leq n/2$ and
$\card{\Gamma'_T(A)} \leq ca$ in the union of $t$ random independent
spanning trees is at most
\begin{align}\label{eq:unionBound}
\sum_{a=1}^{\floor{n/2}} \binom{n}{a} \binom{n}{\floor{ca}} &\Pr( \Gamma_T(A) \subseteq [a+ca] )^t.
\end{align}
We will bound different parts of this sum in two ways: First, for $a \leq n/12$,
we use the random walk construction of a random spanning tree which, as we will see,
can be interpreted as
every vertex in $A$ picking a random neighbor (but not in a completely independent way).
Second, for $a \in (n/12, n/2]$, we look at all the edges of the cut as if they were independent
by means of negative correlation.

So, for the first part of the sum in \eqref{eq:unionBound}, $a \leq n/12$,
consider a random walk on $V$, whose states are denoted $(X_1, X_2, \dotsc)$,
starting outside of $A$, that defines a random spanning tree (as in
the random walk algorithm). Let $\tau_i$ be the first time that the walk
has visited $i$ different vertices of $A$. For $i=1, \dotsc, a-1$,
let $Y_i = \tau_{i+1} - \tau_{i}$ (the gap between first visits $i$ and $i+1$).
We have that the random variables $Y_i$ are
independent. Let $Z_i$ be the indicator of ``$Y_i =1$'', and let $Z
= \sum_{i=1}^{a-1} Z_i$ be the number of adjacent first visits. We
have
\[
\Pr(Z_i = 1) = \Pr(Y_i = 1) = \frac{a-i}{n-1} .
\]
We now give an upper bound to the probability that the predecessor to the first visit
of vertex $i$ in $[a]$ is in $[a+ca]$, given that this predecessor is not a first visit itself
(in this case, the edge coming into $i$ is within $[a+ca])$.
That is, for $2\leq i \leq a$,
\begin{align*}
\Pr(X_{(\tau_i)-1} \in [a+ca] \giventhat Z_{i-1}=0)
    &= \frac{\floor{ca}+i-1}{n-a+i-1} \leq \frac{ca+i-1}{n-a+i-1} \\
\Pr(X_{(\tau_1)-1} \in [a+ca]) &= \frac{\floor{ca}}{n-a} \leq \frac{ca}{n-a}.
\end{align*}
Thus, using edges added when the walk goes from $V \setminus A$ to
$A$ and ignoring edges in the other direction:
\begin{align*}
\Pr( \Gamma_T(A) \subseteq [a+ca] )
    &= \sum_{z=0}^{a-1} \Pr( \Gamma_T(A) \subseteq [a+ca] \giventhat Z = z) \Pr(Z = z) \nonumber\\
    &\leq \sum_{z=0}^{a-1} \left(\prod_{j=1}^{a-z} \frac{a+ca - j}{n - j}\right) \binom{a-1}{z} \left(\prod_{i=1}^{z} \frac{a-i}{n-1}\right) \nonumber \\
    &\leq \sum_{z=0}^{a-1} \binom{a-1}{z} \left( \frac{a+ca}{n}\right)^{a-z} \left( \frac{a}{n}\right)^z \\
    &= \frac{a+ca}{n} \left( \frac{a+ca}{n} + \frac{a}{n}\right)^{a-1} \\
    &\leq \left(\frac{2(1+c)a}{n}\right)^a.
\end{align*}
We now use this in \eqref{eq:unionBound}, for $a \leq n/12$. Let $K= 2(1+c)$.
\begin{align*}
\sum_{a=1}^{\floor{n/12}} \binom{n}{a} \binom{n}{\floor{ca}} &\Pr( \Gamma_T(A) \subseteq [a+ca] )^t \\
    &\leq \sum_{a=1}^{\floor{n/12}} \left(\frac{en}{a}\right)^a \left(\frac{en}{ca}\right)^{ca} \left(\frac{aK}{n}\right)^{at} \\
    &= \sum_{a=1}^{\floor{n/12}} \alpha^a K^{at} \left(\frac{a}{n}\right)^{a(t-1-c)}
\qquad \text{(where $\alpha = \frac{e^{1+c}}{c^c}$)} \\
    &\leq \sum_{a=1}^{\floor{\sqrt{n}}} \alpha^a K^{at} \left(\frac{1}{\sqrt{n}}\right)^{a(t-1-c)} +
        \sum_{a=\floor{\sqrt{n}}+1}^{\floor{n/12}} \alpha^a K^{at} \left(\frac{1}{12}\right)^{a(t-1-c)} \\
    &\leq \left[\alpha K^t n^{-(t-1-c)/2} +
        \left(\frac{\alpha K^t}{12^{t-1-c}}\right)^{\floor{\sqrt{n}}+1} \right] \frac{1}{1-\alpha K^t 12^{-(t-1-c)}}
\end{align*}
which goes to 0 as $n\goesto \infty$ when $\alpha K^t/12^{t-1-c}<1$,
and this happens for $t = 2$ and a sufficiently small constant $c$.

For the rest of the sum in \eqref{eq:unionBound}, $a \in (n/12,n/2]$, we use negative correlation
of the edges of a random spanning tree $T$ (Section \ref{sec:negative_correlation})
to estimate the probability that $\Gamma_T(A) \subseteq [a+ca]$.
Any fixed edge from $K_n$ appears in $T$ with probability $2/n$.
We have that $\Gamma_T(A) \subseteq [a+ca]$ iff no edge between
$A$ and $V \setminus [a+ca]$ is present in $T$, and negative correlation
(Equation \eqref{eqn:negative_correlation_complement})
implies that this happens
with probability at most $(1-2/n)^{a(n-(a+ca))}$. Thus,
\begin{align*}
\sum_{a=\floor{n/12}+1}^{\floor{n/2}} \binom{n}{a} \binom{n}{\floor{ca}} &\Pr( \Gamma_T(A) \subseteq [a+ca] )^t \\
    &\leq \sum_{a=\floor{n/12}+1}^{\floor{n/2}}
        \left(\frac{en}{a}\right)^a
        \left(\frac{en}{ca}\right)^{ca} \left(1-\frac{2}{n}\right)^{ta(n-(a+ca))} \\
    &\leq n \sup_{\gamma \in [1/12,1/2]} \left(\frac{e}{\gamma}\right)^{\gamma n} \left(\frac{e}{c \gamma}\right)^{c \gamma n}
        \left(1-\frac{2}{n}\right)^{t\gamma n(n-(1+c)\gamma n))}\\
    &\leq n \sup_{\gamma \in [1/12,1/2]} \left(\frac{(e/\gamma)^{1+c}}{c^c}\right)^{\gamma n}
        e^{-2t\gamma n(1-(1+c)\gamma)} \\
    &= n \sup_{\gamma \in [1/12,1/2]} \left(\frac{(e/\gamma)^{1+c}}{c^c e^{2 t (1-(1+c)\gamma)}}\right)^{\gamma n}
\end{align*}
For any fixed $c>0$, the function
\[
f(\gamma) = \frac{(e/\gamma)^{1+c}}{c^c e^{2 t (1-(1+c)\gamma)}}
\]
is convex for $\gamma > 0$ and hence the $\sup$ is attained at one of
the boundary points $1/12$ and $1/2$, and the function is strictly less than 1 at these boundary points for $t=2$
and a sufficiently small constant $c$. This implies that this sum goes to 0 as $n\goesto \infty$.
\end{proof}

\section{Expansion when base graph is a bounded-degree graph:
positive result}

In this section we consider graphs with bounded degrees. To simplify
the presentation we restrict ourselves to regular graphs; it is easy
to drop this restriction at the cost of extra notation. We show that
for constant degree graphs the edge expansion is captured fairly
well by the union of a small number of random spanning trees.

\begin{proof}[Proof (of Theorem~\ref{thm:constant:positive}).]
It follows by the random walk construction of random spanning trees
that for any edge $(u,v) \in E$ we have $\Pr[(u,v) \in T] \geq 1/d(u).$
To see this, note that if we start the random walk at vertex $u$
then with probability $1/d(u)$ the first traversed edge is $(u,v)$,
which then gets included in $T$.  Thus for $A \subset V$, we have
that
$$\E[|\delta_{T_G}(A)|] \geq \frac{1}{d} \cdot |\delta_{G}(A)|.$$

We would now like to use the above expectation result to prove our
theorem.  Recall the definition of random variables $X_e$ from
Section~\ref{sec:negative_correlation}: For edge $e \in E$, $X_e$ is
the indicator random variable taking value $1$ if $e \in T$, and
value $0$ otherwise. Thus we have $|\delta_T(A)| = \sum_{e \in
\delta_G(A)} X_e.$ We want to show that $\sum_{e \in \delta_G(A)}
X_e$ is not much smaller than its expectation with high probability.
Random variables $X_e$ are not independent.  Fortunately, they are
negatively correlated as we saw in
Section~\ref{sec:negative_correlation}, which allows us to use
Theorem~\ref{thm:Chernoff_negative}:
\begin{align} \label{eqn:Chernoff_tree}
\Pr\left[\sum_{e \in \delta_G(A)} X_e < p |\delta_G(A)| - \lambda\right]
    < e^{-\lambda^2/(2 p |\delta_G(A)|)}
    \leq e^{-\lambda^2 /(2 |\delta_G(A)|)},
\end{align}
where $p$ is the average of $\Pr[X_e = 1]$ for $e \in \delta_G(A)$.
Since $\Pr[X_e = 1] \geq 1/d$ for all edges $e$, we have  $p \geq
1/d$, and for $\lambda = (p-1/(2d))|\delta_G(A)|$ we have
\begin{align*}
\Pr[|\delta_{T_G}(A)| < \frac{1}{2d} |\delta_G(A)|]
    < e^{-\frac{|\delta_G(A)|}{8d^2}}.
\end{align*}

Which gives
\begin{align} \label{eq:badevent}
\Pr[|\delta_{U^k_G}(A)| < \frac{1}{2d} |\delta_G(A)|]
< e^{-\frac{k |\delta_G(A)|}{8d^2}}.
\end{align}

Now we estimate the probability that there is a \emph{bad cut},
namely a cut $A$ such that $|\delta_{U^k_G}(A)| = a$ and
$|\delta_G(A)| \geq \alpha a \ln{n}$. To do this we first look at
cuts of size $a$ in the first random tree, which have size at least
$\alpha a\ln{n}$ in $G$ (This step is necessary: the modified
Chernoff bound that we use is only as strong as the independent
case, and when edges are chosen independently one is likely to get
isolated vertices; looking at the first tree ensures that this does
not happen). In order to be bad, these cuts have to have small size
in all the remaining trees. The probability of that happening is
given by \eqref{eq:badevent}. The number of cuts in the first tree
of size $a$ is clearly no more than $ \binom{n-1}{a} <
\binom{n}{a}$, as there are $\binom{n-1}{a}$ ways of picking $a$
edges out of $n-1$, although not all of these may correspond to
valid cuts. Then, the probability that a bad cut exists is at most
\begin{align*}
\sum_{a=1}^{n/\ln{n}} \binom{n}{a} e^{-\frac{(k-1) \alpha a \ln{n}}{8d^2}}
&\leq \sum_{a=1}^{n/\ln{n}} \left(\frac{en}{a}\right)^a e^{-
  \frac{(k-1)\alpha a\ln{n}}{8d^2}}  \\
&= \sum_{a=1}^{n/\ln{n}} \exp\left(\Bigl(\ln(en/a) -
\frac{(k-1) \alpha \ln{n}}{8d^2}\Bigr)\; a\right) \\
&= \sum_{a=1}^{n/\ln{n}} \exp\left( \biggl(\ln(e/a)+ \Bigl(1 - \frac{(k-1)\alpha}{8d^2}\Bigr)\ln{n} \biggr)\; a \right).
\end{align*}
Choosing $(k-1)\alpha > 9d^2$ makes the above sum $o(1)$.
\end{proof}


\section{Expansion when base graph is a bounded-degree graph: negative result}
Here we show that Theorem~\ref{thm:constant:positive} is best
possible up to a constant factor for expansion:

\begin{proof}[Proof (of Theorem~\ref{thm:constant:negative}).]
We begin with a $d$-regular edge expander $G'$ on $n$ vertices with
a Hamiltonian cycle (such graphs are known to exist), where $d>2$ is
a fixed integer. Let $0 < \ell < \log{n}$ be an integer to be chosen
later, and let $H$ be a Hamiltonian path in $G'$. Subdivide $H$ into
subpaths $P_1, \ldots, P_{n/\ell}$ each of length $\ell$ (to keep
the formulas simple we suppress the integrality issues here which
are easily taken care of).

For two subpaths $P_i$ and $P_j$, we say that they \emph{interact}
if $(P_i \cup \Gamma'(P_i)) \cap (P_j \cup \Gamma'(P_j)) \neq
\emptyset $. Since $G'$ is $d$-regular, $|\Gamma'(P_i)| \leq d
\ell$.  So, any subpath can interact with at most $d^2\ell$ other
subpaths (this bound is slightly loose). Thus we can find a set $I$
of $\frac{1}{d^2 \ell} \cdot n/\ell$ paths among $P_1, \ldots,
P_{n/\ell}$, so that no two paths in $I$ interact.

We now describe the construction of $G$, which will be obtained by
adding edges to $G'$. For each path $P \in I$, we do the following.
Add an edge between the two end-points of $P_i$, if such an edge did
not already exist in $G'$. If the subgraph $G[\Gamma'(P_i)]$ induced
by the neighborhood of path $P_i$ does not have a Hamiltonian cycle,
then we add edges to it so that it becomes Hamiltonian. Clearly, in
doing so we only need to increase the degree of each vertex by at
most $2$. The final graph that we are left with is our $G$. For each
path $P \in I$ we fix a Hamiltonian cycle in $G[\Gamma'(P)]$, and we
also have the cycle of which $P$ is a part. We denote these two
cycles by $C_1(P)$ and $C_2(P)$.

We will generate a random spanning tree $T$ of $G$ by the random
walk algorithm starting the random walk at some vertex outside of
all paths in $I$. For $P \in I$, we say that event $E_P$ (over the
choice of a random spanning tree $T$ of $G$) occurs if the random
walk, on first visit to $ C_1(P) \cup C_2(P) $, first goes around
$C_1(P)$ without going out or visiting any vertex twice, and then it
goes on to traverse $C_2(P)$, again without going out or visiting
any vertex twice until it has visited all vertices in $C_2(P)$. For
all $P \in I$ we have
\begin{align}
\Pr[E_P] \geq 1/(d+2)^{|C_1(P)|+|C_2(P)|-1} \geq 1/(d+2)^{(d+1)\ell - 1}.
\end{align}
If event $E_P$ happens then in the resulting tree $T$ we have
$|\delta_T(V(P))|=1$. Thus our goal will be to show that with
substantial probability there is a $P \in I$ such that $E_P$
happens. Since no two paths in $I$ interact with each other, events
$E_P$ are mutually independent. If we are choosing $k$ random
spanning trees, then define $E^k_P$ to be the event that $E_P$
occurs for all $k$ spanning trees. Clearly, $\Pr[E^k_P]=\Pr[E_P]^k$.
Then the probability that $E^k_P$ doesn't occur for any $P \in I$ is
at most
\begin{align*}
\left(1 -  \frac{1}{(d+2)^{k(d+1)\ell - k}}\right)^{|I|}
    &= \left(1 -  \frac{1}{(d+2)^{k(d+1)\ell -k}}\right)^{\frac{n}{d^2 \ell^2}} \\
    &\leq \exp\left(-\frac{n}{(d+2)^{k(d+1)\ell - k + 2}\ell^2}\right).
\end{align*}

It follows readily that there is a constant $C$ (that depends on
$d$) such that for $\ell k \leq C\log{n}$ the above probability is
$o(1)$. Hence, with probability $1-o(1)$ there is a path $P \in I$
such that $|\delta_{U^k_G}(V(P))| \leq k$.  The edge expansion of
$P$ therefore is $k/\ell= k^2/(C\log{n})$ for $\ell=C(\log{n})/k$.
\end{proof}

\section{Splicers of random graphs}

We will show a random process on random graphs that generates random
spanning trees with a distribution that is very close to the
uniform distribution on the complete graph. The process first directs
edges to mimic the distribution of a directed random graph.

Given an undirected graph $H$ and a parameter $0 < p \leq 1$,
construct a random directed graph denoted $D_p(H)$ with vertex set
$V(H)$ and independently for every edge $(u,v)$ of $H$:
\begin{itemize}
    \item edges $(u,v)$ and $(v,u)$ with probability
        $\frac{-p-2 \sqrt{1-p}+2}{p}$,
    \item only edge $(u,v)$ with probability
        $\frac{p+\sqrt{1-p}-1}{p}$, and
    \item only edge $(v,u)$ with probability
        $\frac{p+\sqrt{1-p}-1}{p}$.
\end{itemize}
If $H$ is random according to $G_{n,p}$, then $D_p(H)$ is random
with each edge picked with probability $q=1-\sqrt{1-p}$. Note that
$p/2 \leq q \leq p$.

Let $T$ be the uniform distribution on spanning trees of
$K_n$. We now describe Process $B_p$, which is a random
process that given an undirected graph $H$ and a parameter $0 < p
\leq 1$ generates a spanning tree with a distribution that we denote
$T_{p,H}$ Consider the following random process that generates a
walk in $D_p(H)$ or stops with no output:
\begin{enumerate}
\item Start at a vertex $v_0$ of $D_p(H)$.
\item At a vertex $v$, an edge is traversed as follows. Suppose
    $d_1(v)$ out of $d(v)$ outgoing edges at $v$ are previously
    traversed. Then, the probability of picking a previously
    traversed edge is $1/(n-1)$ while the probability for each
    new edge is
    \[
        \frac{1-\frac{d_1(v)}{n-1}}{d(v)-d_1(v)}.
    \]
\item If all vertices have been visited, output the walk and
    stop. If this has not happened and at the current vertex $v$
    one has $d_1(v) = d(v)$, stop with no output.
\end{enumerate}
As in the random walk algorithm, the spanning tree given by Process $B_p$
(if it succeeds in visiting all the vertices) is the set of edges
that are used on first visits to each vertex, but the random
sequence of edges is different here.

A covering path of a graph is a path passing through all vertices.
Let $D$ be the distribution on covering paths of the (undirected)
complete graph starting at a vertex $v_0$ where a random path is
generated by a random walk that starts at $v_0$ and walks until it
has visited all the vertices. Let $D_p$ be the distribution on
covering paths of the complete graph given by first choosing $H$
according to $G_{n,p}$ and running Process $B_p$ starting from
$v_0$.

\begin{lemma}\label{lem:pathcoupling}
There exists an absolute constant $c$ such that for $p > c\log n/n$
the total variation distance between the distributions $D$ and $D_p$
is $o(1)$.
\end{lemma}
\begin{proof}
We will couple $D$ and $D_p$ so that the walk in $D$ picks the same
edges as the walk in $D_p$, but if $D_p$ fails, then $D$ continues
its random walk. Then this covering walks coincide whenever $D_p$
succeeds, and thus the probability of success is an upper bound to
the total variation distance between $D$ and $D_p$. Now, $D_p$ does
not fail if every vertex in $H_d$ has out-degree at least $c_1 \log
n$ and Process $B_p$ does not visit any vertex more than $c_2 \log
n$ times, for $c_1 > c_2$. A Chernoff bound gives $c$ and $c_1$ such
that the first part happens with probability $1-o(1)$. For the
second part, we observe that if there is no failure then Procedure B
behaves exactly like a random walk in the complete graph, and
therefore it visits all vertices in at most $ c_3 n \log n $ steps
with probability $1-o(1)$ for some constant $c_3$ (this is
essentially the coupon collector's problem with $n-1$ coupons, see
\cite[Section 3.6 and Chapter 6]{RandomizedAlgorithms}) and a walk
of that length does not visit any vertex more than $ c_2 \log n$
times with probability $1-o(1)$ for some constant $c_2$ (by a
straightforward variation of the occupancy problem in \cite[Section
3.1]{RandomizedAlgorithms}). \additionaldetails{The variation: the
probability that a fixed vertex is visited exactly $i$ times in a
walk of length $(n-1) \log n$ is at most
\[
\binom{(n-1)\log n}{i} (\frac{1}{n-1})^i \leq (\frac{e \log n}{i})^i.
\]
Thus, the probability that our fixed vertex is visited more than $k$
times is at most
\begin{align*}
\sum_{i=k}^{(n-1)\log n}
    &\leq (\frac{e \log n}{k})^k (1+\frac{e \log k}{k} + (\frac{e \log k}{k})^2 + \dotsb) \\
    &= (\frac{e \log k}{k})^k \frac{1}{1-e (\log n) /k}.
\end{align*}
For $k = e^2 \log n$ this is less than $1/n^2$. A union bound over
all vertices tells us that no vertex is visited more than $e^2 \log
n$ times with probability $1-o(1)$. (similar to \cite[Section
3.1]{RandomizedAlgorithms})}
\end{proof}

Let $T_p$ be the distribution on trees obtained by first choosing
$H$ from $G_{n,p}$ and then generating a random spanning tree
according to Process $B_p$.
\begin{lemma}\label{lem:coupling}
There exists an absolute constant $c$ such that for $p > c\log n/n$
the total variation distance between the distributions $T$ and $T_p$
is $o(1)$.
\end{lemma}
\begin{proof}
This is immediate from Lemma \ref{lem:pathcoupling}, as random trees
from  $T$ or $T_p$ are just functions of walks from $D$ or $D_p$,
respectively.
\end{proof}

\begin{proof}[Proof (of Theorem \ref{thm:expander_random}).]
In the random graph $H$, we generate two random trees by using one
long sequence of edges, with a breakpoint whenever we complete the
generation of a spanning tree. In the complete graph also, we
generate two trees from such a sequence obtained from the uniform
random walk. Using the same coupling as in Lemma \ref{lem:coupling}
we see that these distributions on these sequences have variation
distance $o(1)$. Therefore the spanning trees of $H$ obtained by the
first process have total variation distance $o(1)$ to random
spanning trees of the complete graph. By Theorem
\ref{thm:completegraph}, the union of these trees has constant
expansion with probability $1-o(1)$ overall.
\end{proof}

With this results we are ready to prove our theorem about sparsifiers of random graphs:
\begin{proof}[Proof (of Theorem \ref{thm:sparsifier_gnp}).]
We need the fact that for sufficiently large constant $C$, with probability
$1-o(1)$, all cuts $\delta_H(A)$ in random graph $H$ satisfy
$$ c_3 p|A|(n-|A|) \leq |\delta_H(A)| \leq c_4 p|A|(n-|A|).$$
This is well-known and follows
immediately from appropriate Chernoff-type bounds.

We only need to prove the theorem for $|A| \leq n/2$.
We now prove the first inequality in the statement of the theorem.
By Theorem~\ref{thm:completegraph}, with probability $1-o(1)$, for any
$A \subset V$ such that $|A| \leq n/2$, we have $|\delta_{H'}(A)| \geq c_5 |A|$,
and so $w(\delta_{H'}(A) \geq c_5 |A| pn \geq c_5 p |A|(n-|A|) \geq
\frac{c_5}{c_4} |\delta_H(A)|$.

For the second inequality in the statement of the theorem, we need the fact that
the maximum degree of a vertex in a random spanning tree in the complete graph
is $O(\log{n})$.  So the same holds for random spanning trees generated by
process $B_p$.  We then have $|\delta{H'}(A)| \leq c_6 \log{n} |A|$, and so
$w(\delta_{H'}(A)) \leq c_6 \log{n} |A| p n \leq
\frac{2c_6}{c_2} \log{n} |\delta_{H'}(A)|$.
\end{proof}

\section{Discussion}
The problem of scalable routing in the presence of failures has
motivated a novel construction of sparse expanders. The use of trees
is particularly natural for routing. Our results suggest using a
constant number of trees in total for routing, as opposed to the
norm of one or more trees per destination. Further, the manner in
which the trees are obtained is simple to implement and can lead to
faster recovery since (a) paths exist after several failures and (b)
fewer trees need to be recomputed in any case.

One aspect of splicers that we have not fully explored is the
stretch of the metric induced by them. For the case of the complete
graph, it is not hard to see that the diameter is $O(\log n)$ and
hence so is the expected stretch for a pair of random vertices. This
continues to hold for $G_{n,p}$, in fact giving better bounds for
small $p$ (expected stretch of $O(\log \log n)$ for $p=\mbox{
poly}(\log n)/n$). It remains to study the stretch of splicers for
arbitrary graphs or bounded-degree graphs. This seems to be an
interesting question since on the complete graph, the expected
stretch on one tree is $\Theta(\sqrt{n})$ while that of two trees is
$O(\log n)$.

Finally, Process $B_p$ appears interesting to study on its own.

\bibliographystyle{abbrv}    
\bibliography{btrees}

\end{document}